\documentclass[twocolumn,showpacs,preprintnumbers,amsmath,amssymb,notitlepage]{revtex4-1}

\usepackage{amsfonts}
\usepackage{amsmath}
\usepackage{amssymb}
\usepackage{graphicx}
\usepackage{dcolumn}
\usepackage{bm}
\usepackage{tabularx}
\usepackage{epstopdf}
\usepackage{xcolor}
\usepackage{mathrsfs}
\usepackage{mathtools}
\usepackage{float}
\usepackage{algpseudocode}
\usepackage{verbatim}
\usepackage{footnote}
\usepackage{url}

\begin{document}

\title{Detecting Core-Periphery Structures by Surprise}

\author{Jeroen van Lidth de Jeude}
\affiliation{IMT School for Advanced Studies, Piazza S.Francesco 19, 55100 Lucca - Italy}
\author{Guido Caldarelli}
\affiliation{IMT School for Advanced Studies, Piazza S.Francesco 19, 55100 Lucca - Italy}
\author{Tiziano Squartini}
\affiliation{IMT School for Advanced Studies, Piazza S.Francesco 19, 55100 Lucca - Italy}
\date{\today}

\begin{abstract}
Detecting the presence of mesoscale structures in complex networks is of primary importance. This is especially true for financial networks, whose structural organization deeply affects their resilience to events like default cascades, shocks propagation, etc. Several methods have been proposed, so far, to detect \emph{communities}, i.e. groups of nodes whose internal connectivity is significantly large. Communities, however do not represent the only kind of mesoscale structures characterizing real-world networks: other examples are provided by bow-tie structures, core-periphery structures and bipartite structures. Here we propose a novel method to detect statistically-significant \emph{bimodular} structures, i.e. either bipartite or core-periphery ones. It is based on a modification of the \emph{surprise}, recently proposed for detecting communities. Our variant allows for bimodular nodes partitions to be revealed, by letting links to be placed either 1) within the core part and between the core and the periphery parts or 2) between the layers of a bipartite network. From a technical point of view, this is achieved by employing a multinomial hypergeometric distribution instead of the traditional, binomial hypergeometric one; as in the latter case, this allows a p-value to be assigned to any given (bi)partition of the nodes. To illustrate the performance of our method, we report the results of its application to several real-world networks, including social, economic and financial ones.
\end{abstract}

\maketitle

\section*{Introduction}

Detecting the presence of mesoscale structures in complex networks is of primary importance \cite{Fortunato2016,Bisma2017}. This is especially true for financial networks, whose structural organization deeply affects their resilience to shocks propagation, node failures, etc. \cite{Craig2010,IntVeld2014,Fricke2014,Luu2018}. Several methods have been proposed, so far, to detect communities, i.e. groups of nodes whose ``internal'' connectivity is significantly large. Communities, however, do not represent the only kind of mesoscale structures characterizing real-world networks: other examples are provided by bow-tie, core-periphery and bipartite structures. In what follows, we will focus on the last two types of topological structures.

The intuitive notion of core-periphery network, as a configuration consisting of a densely-connected bunch of nodes (i.e. the core) and low-degree nodes preferentially connected to the core (i.e. the periphery ones) has been firstly formalized by Borgatti \& Everett: in \cite{Borgatti2000} a score function indicating the extent to which a given graph partition deviates from an ideal core-periphery configuration (where the core is \emph{fully} connected and the peripherical nodes are \emph{only} linked to the core ones) was defined. Several later works adopted the same approach \cite{IntVeld2014,Fricke2014,Boyd2006}, accompanying the error score with a significance level, computed on a properly-generated ensemble of networks (see \cite{Csermely2013} for a review on the topic). Detection of bipartitiveness has been approached similarly, by quantifying the deviation of an observed graph partition from the ideal bipartite configuration (where edges exist only between layers and not within them) \cite{Holme2003,Estrada2005}. 

Conversely, in recent years the detection of mesoscale structures has been faced by adopting a bottom-up approach, i.e. by defining a benchmark model against which to compare the actual network structure: in \cite{Zhang2014} the authors aim at identifying the most likely generative model that may have produced a given partition; in \cite{Barucca2015,Barucca2018} the authors compare the likelihood values of a Stochastic Block Model tuned to reproduce either a core-periphery or a bipartite structure; similarly, in \cite{Kojaku2017} the authors adopt a Random Graph Model to find multiple core-periphery pairs in networks and in \cite{Kojaku2018} the same authors employ the Configuration Model as a benchmark, showing that a single core-periphery structure can never be significant under it, seemingly confirming recent findings by the authors of the present paper \cite{IntVeld2014,DeJeude2018}.

We contribute to this stream of research by proposing a novel method to detect statistically-significant bimodular structures (i.e. either bipartite or core-periphery ones). To this aim, we build upon the results of the papers \cite{Aldecoa2013,Aldecoa2013b,Nicolini2016} and on the very last comment that can be found in \cite{Traag2015}, by adopting a surprise-like score function. Our choice is dictated by the versatility of this kind of quantity that allows us to consider undirected as well as directed (binary) networks, a desirable feature that many of the aforementioned algorithms do not have.

The paper is organized as follows: the Methods section is devoted to illustrate the definition of our bimodular surprise; the results of its application to real-world networks are shown in the Results section and further discussed in the Discussion section where future perspectives are also presented.

\begin{figure*}[t!]
\centering
\includegraphics[width=0.5\textwidth]{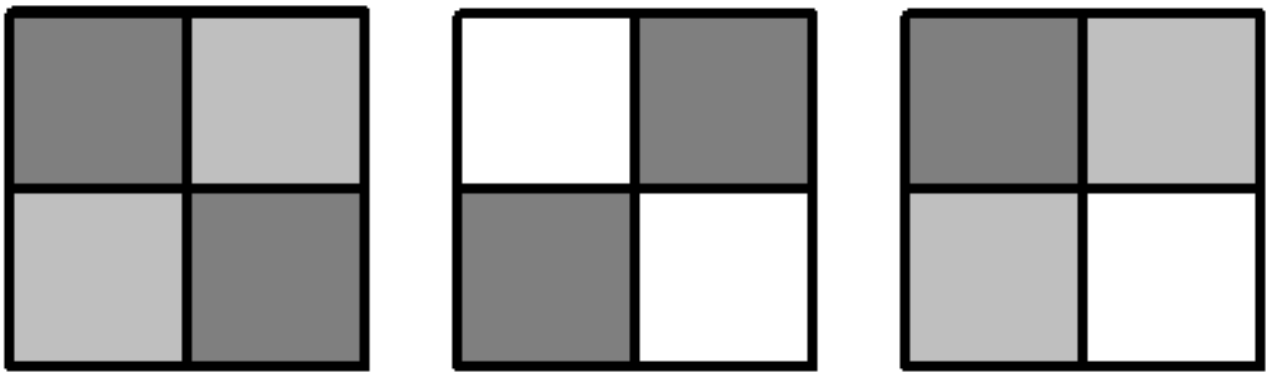}
\caption{Examples of mesoscale network structures: a traditional community structure is shown on the left, a purely bipartite network is shown in the middle and a core-periphery structure is shown on the right. White blocks represents subsets of nodes whose link density is zero, darker blocks represents subsets of nodes whose link density is higher.}
\label{fig1}
\end{figure*}

\section*{Methods}

Let us first discuss the limitations of traditional surprise whenever employed to detect bimodular structures. In what follows we will implement the following definition of surprise \cite{Aldecoa2013,Aldecoa2013b,Nicolini2016}

\begin{equation}
S\equiv\sum_{i\geq l^*}\frac{\binom{V_{int}}{i}\binom{V-V_{int}}{L-i}}{\binom{V}{L}};
\label{eq1}
\end{equation}
the sum runs up to the value $i=\min\{L,V_{int}\}$, where $V$ is the volume of the network, coinciding with the total number of nodes pairs (i.e. $V=\frac{N(N-1)}{2}$ in the undirected case and $V=N(N-1)$ in the directed case), $V_{int}$ is the total number of intracluster pairs (i.e. the number of nodes pairs \emph{within} the individuated communities), $L$ is the total number of links and $l^*$ is the observed number of intracluster links (i.e. \emph{within} the individuated communities.

The hypergeometric distribution describes the probability of observing $i$ successes in $L$ draws (without replacement) from a finite population of size $V$ that contains exactly $V_{int}$ objects with the desired feature (in our case, being an intracluster pair), each draw being either a success or a failure: surprise is the p-value of such a distribution, testing the statistical significance of the observed partition against the null hypothesis that the intracluster link density $p_{int}=\frac{l^*}{V_{int}}$ is compatible with the density $p=\frac{L}{V}$ characterizing the (Directed) Random Graph Model.

\subsection*{The limitations of surprise}

While traditional surprise $S$ is suited for community detection, it suffers from several limitations whenever employed to detect bimodular mesoscale structures.

\paragraph*{Bipartite networks.} Let us first consider a purely bipartite, undirected network, as the one shown in fig. \ref{fig1}, whose first and second layer consist of $N_1$ and $N_2$ nodes respectively. Since we would like $S$ to reveal two (empty) communities, we would be tempted to instantiate eq. \ref{eq1} with the values $V=\frac{(N_1+N_2)(N_1+N_2-1)}{2}$, $V_{int}=\frac{N_1(N_1-1)}{2}+\frac{N_2(N_2-1)}{2}$ and $l^*=0$; upon considering, however, that $L\leq V_{int}$, the explicit computation of $S$ reveals that $S=1$ (as follows from the Vandermonde identity). Since $S$ is nothing else than a p-value, a significant partition is expected to satisfy $S\leq S_{th}$, with $S_{th}$ usually chosen to attain the value $0.01$ or $0.05$. In our case, however, the opposite result is obtained: the considered (bi)partition \emph{cannot} be significant, independently from the actual number of connections characterizing the considered configuration. This example highlights one of the limitations of the definition provided in eq. \ref{eq1}.

\begin{figure*}[t!]
\centering
\includegraphics[width=0.25\textwidth]{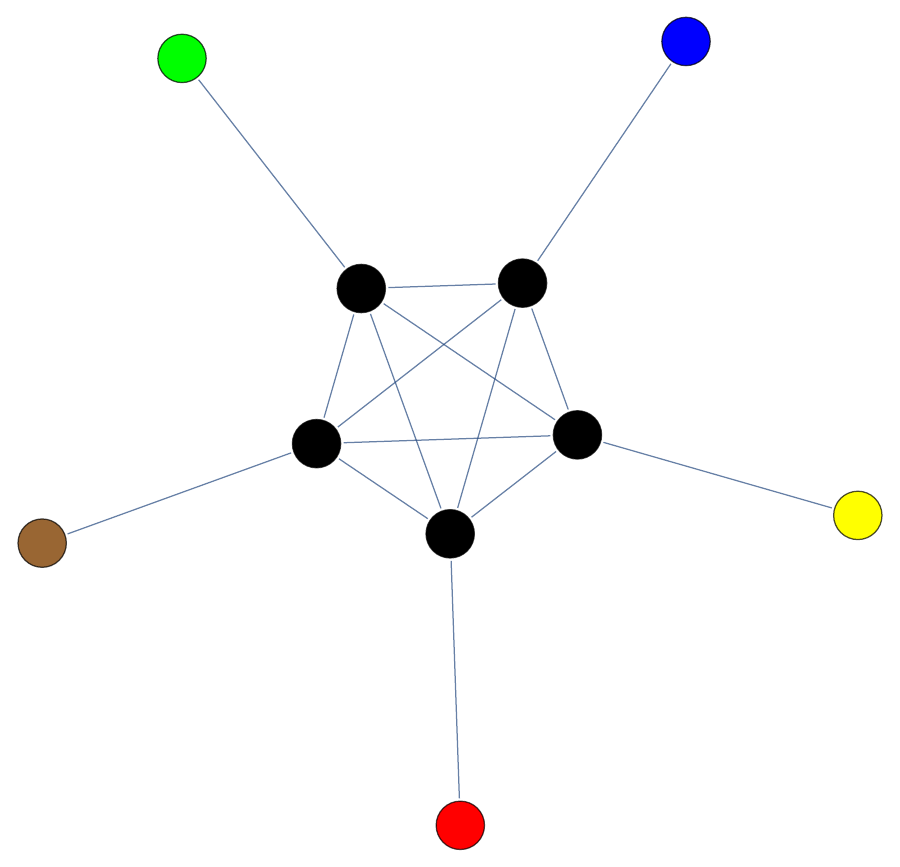}\hspace{3mm}
\includegraphics[width=0.25\textwidth]{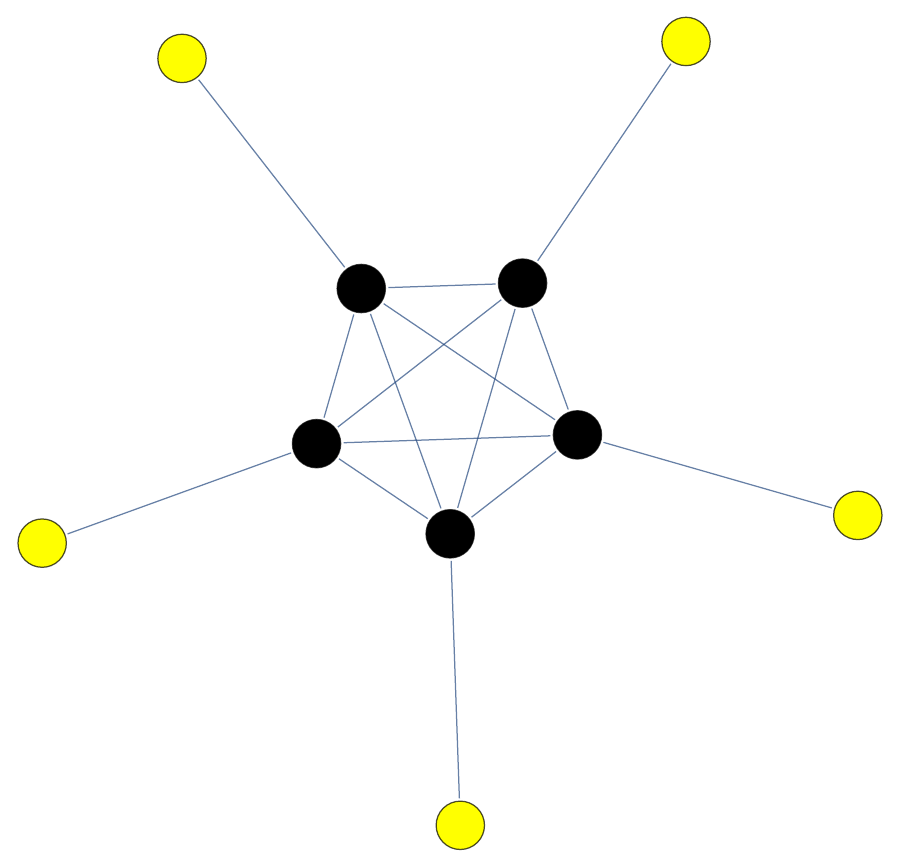}\hspace{3mm}
\includegraphics[width=0.35\textwidth]{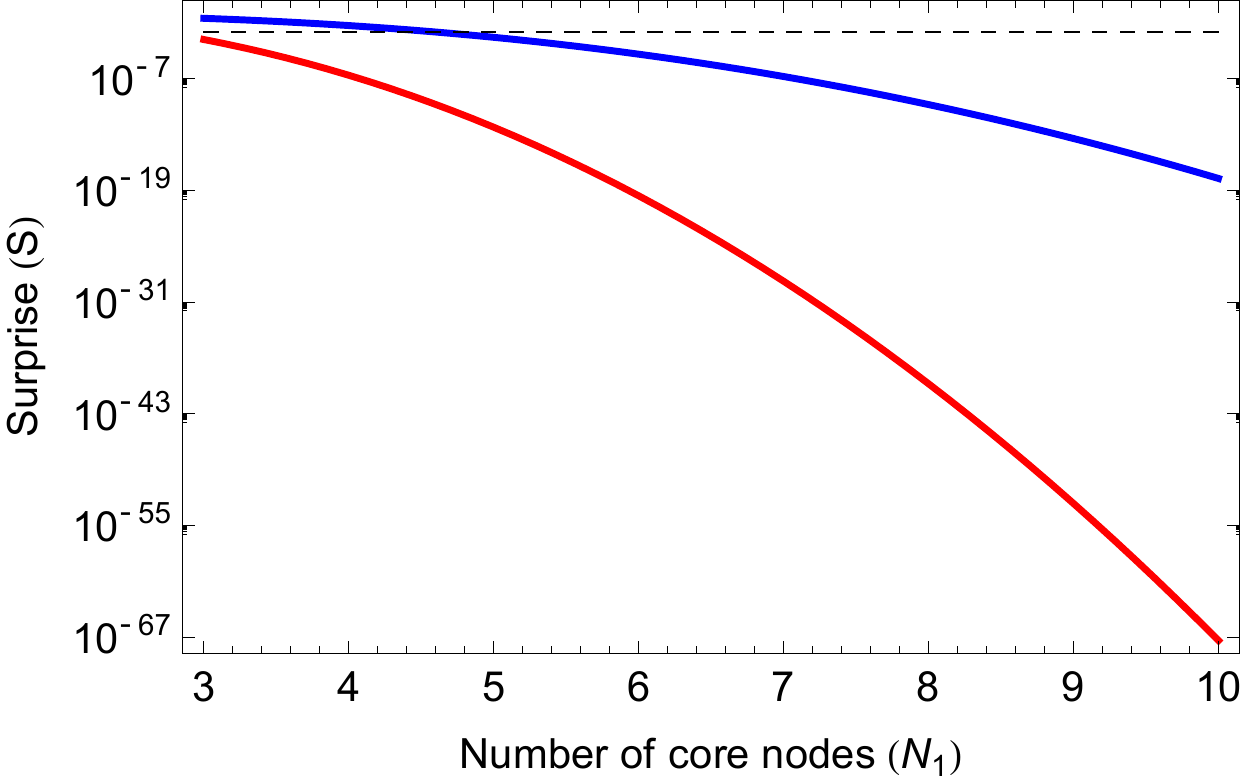}
\caption{Right panel: traditional surprise computed for the two partitions shown in the left and middle panels. The red line refers to the partition constituted by 6 communities (left panel), the blue line referes to the partition constituted by 2 communities (middle panel) and the black, dashed line corresponds to the value $S_{th}=0.05$. As the number of core nodes is risen, both partitions become increasingly significant; the former, however, is always more significant than the latter. The network configuration shown in the left panel is, in fact, recognized as the optimal one, as further confirmed by running the PACO algorithm \cite{Nicolini2016}.}
\label{fig2}
\end{figure*}

\paragraph*{Star-like networks.} Let us now consider proper core-periphery networks: according to the intuitive definition provided in \cite{Borgatti2000}, such configurations are characterized by a densely-connected portion, i.e. the core (in the ideal case $c_{c}\simeq1$) and a sparsely-connected portion, i.e. the periphery (in the ideal case $c_{p}\simeq0$). The density of the intermediate portion is variable, although the chain of inequalities $c_{p}\leq c_{cp}\leq c_{c}$ is always assumed to hold. Let us consider a peculiar example of this kind of networks, i.e. an undirected configuration with a fully connected core plus a periphery of nodes, each of which is connected to just one core node. For the moment, let us suppose that the number of core nodes coincides with the number of periphery nodes and let us instantiate $S$ on a partition that identifies each periphery node as a community on its own while considering the core as a traditional community (see fig. \ref{fig2}). If we consider a core portion of $N_1$ nodes and $N_2=N_1$ peripherical nodes, we have $V=\frac{(N_1+N_2)(N_1+N_2-1)}{2}$, $V_{int}=\frac{N_1(N_1-1)}{2}$, $L=\frac{N_1(N_1-1)}{2}+N_1$ and $l^*=\frac{N_1(N_1-1)}{2}$. Notice that, in this case, only the addendum corresponding to the value $i=l^*=V_{int}$ survives, i.e.

\begin{equation}
S=\frac{\binom{N_1(3N_1-1)/2}{N_1}}{\binom{N_1(2N_1-1)}{N_1(N_1+1)/2}}
\end{equation}
which is of the order of $10^{-2}$ for $N_1=3$ and rapidly decreases as $N_1$ grows (see fig. \ref{fig2}). Since $S<S_{th}=0.05$, such a partition is recovered as significant. As confirmed by running the PACO algorithm \cite{Nicolini2016}, such a configuration - constituted by an unreasonably large number of single-nodes communities - is indeed recognized as the optimal one.

For the sake of comparison, let us calculate $S$ for the ``reasonable'' partition identifying the core and the periphery as two separate communities: in this case, $V=\frac{(N_1+N_2)(N_1+N_2-1)}{2}$, $V_{int}=N_1(N_1-1)$, $L=\frac{N_1(N_1-1)}{2}+N_1$ and $l^*=\frac{N_1(N_1-1)}{2}$. As our explicit calculation reveals, such a partition can indeed be significant but it is not the optimal one (see also fig. \ref{fig2}).

\paragraph*{k-star networks.} Let us now generalize the star-like network model, by considering a graph with $k$ peripherical nodes linked to each core node (see fig. \ref{fig3}). Instantiating $S$ by considering each group of $k$ leaves as a community on its own leads to

\begin{equation}
S=\sum_{i=l^*}^L\frac{\binom{V_{int}}{i}\binom{V-V_{int}}{L-i}}{\binom{V}{L}}
\label{eq:sks}
\end{equation}
with $V=\frac{(N_1+kN_1)(N_1+kN_1-1)}{2}$, $V_{int}=\frac{N_1(N_1-1)}{2}+\frac{N_1k(k-1)}{2}$, $L=\frac{N_1(N_1-1)}{2}+kN_1$ and $l^*=\frac{N_1(N_1-1)}{2}$ (as long as $k\geq 2$, in fact, $L\leq V_{int}$). The expression defined by eq. \ref{eq:sks} is significant only under certain conditions: in particular, {\it a}) for a given $N_1$ value, as $k$ grows surprise becomes increasingly non-significant; {\it b}) for a given $k$ value, as $N_1$ grows surprise becomes increasingly significant. Since the $k$ nodes linked to each core node should be always considered as \emph{non} constituting separate communities, irrespectively from the value of $k$, the findings above point out another detectability limit of surprise that, for certain values of the parameters, misinterprets the (planted) partition under analysis.

\subsection*{A bimodular surprise}

The previous examples have shown that traditional surprise may suffer from some limitations whenever employed to detect bimodular structures. Here we address such an issue by introducing a variant of traditional surprise, designed to detect bimodular mesoscale structures.

Whenever community detection is carried out by maximizing the surprise, links are understood as belonging to \emph{two} different categories, i.e. the \emph{internal} ones (the ones \emph{within} clusters) and the \emph{external} ones (the ones \emph{between} clusters). On the other hand, whenever one is interested in detecting bimodular structures (be they bipartite or core-periphery), \emph{three} different ``species'' of links are needed (e.g. core, core-periphery and periphery links). This is the reason why we need to consider the multinomial version of the surprise, whose definition reads

\begin{figure*}[t!]
\centering
\includegraphics[width=0.25\textwidth]{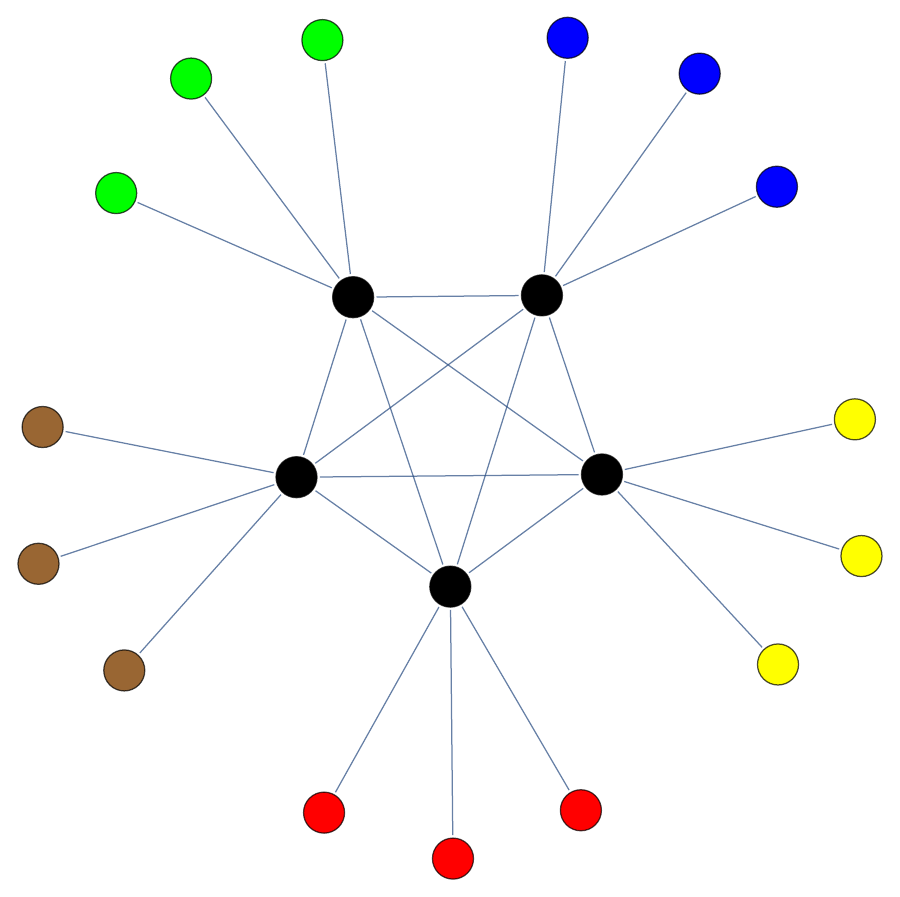}\hspace{3mm}
\includegraphics[width=0.35\textwidth]{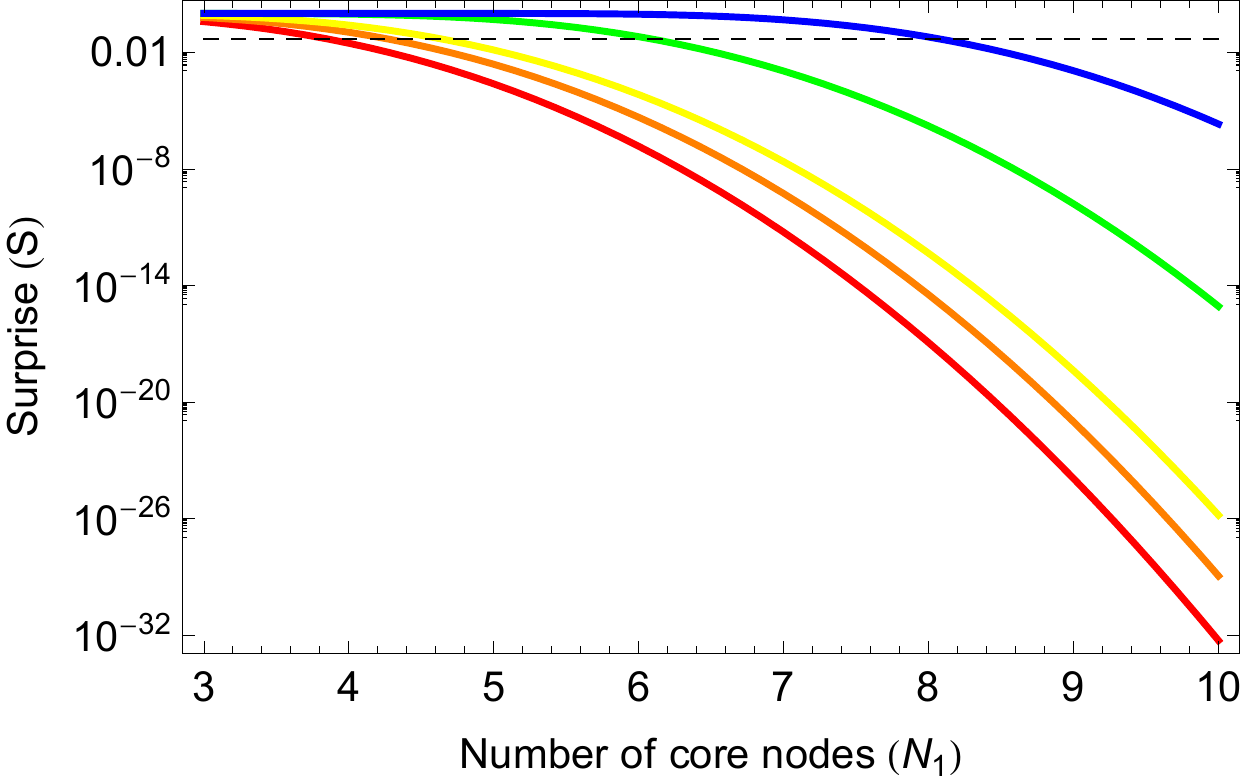}\hspace{3mm}
\includegraphics[width=0.35\textwidth]{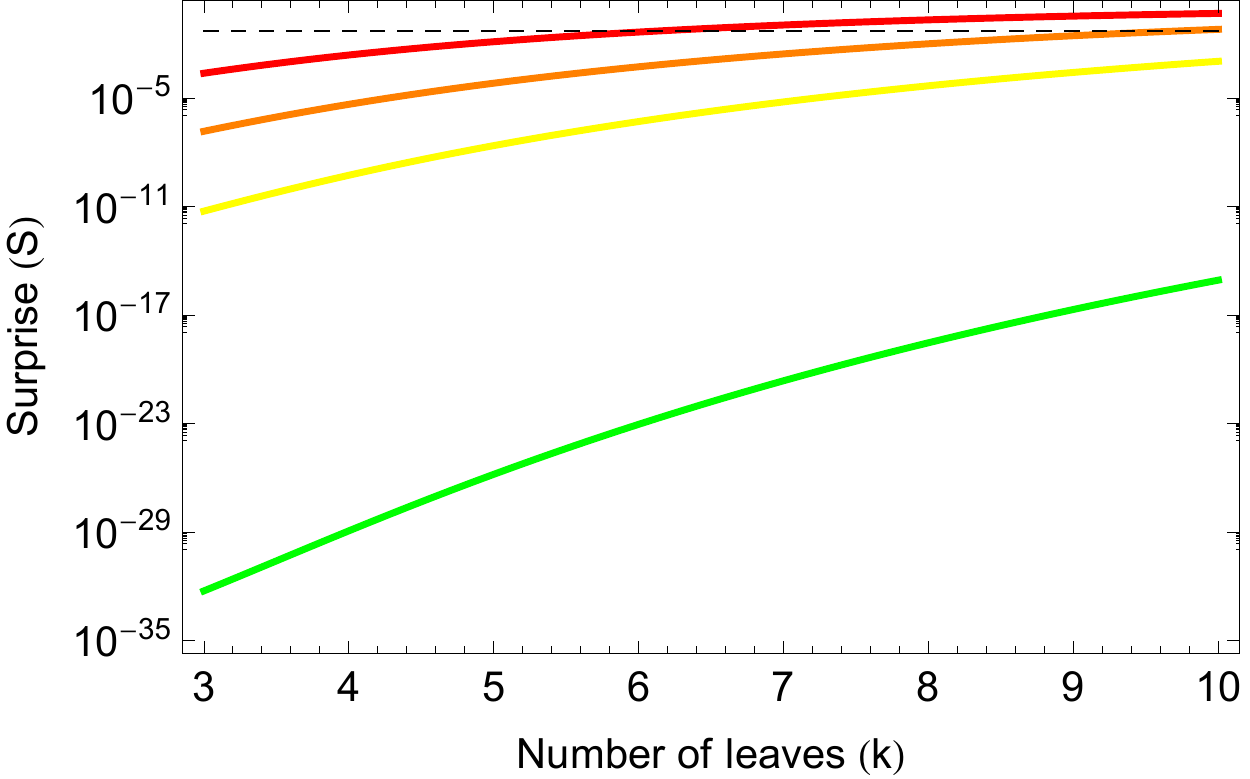}
\caption{Traditional surprise optimization on a k-star network would lead to identify each group of peripherical nodes as a community on its own, although the intracluster density is zero. More precisely, for a given number of leaves ($k=3,4,5,10,20$ as indicated by the red, orange, yellow, green, blue line respectively - middle panel), as the number of core nodes rises, surprise is found to be increasingly significant. Consistently, for a given number of core nodes ($N_1=5,6,7,10$ as indicated by the red, orange, yellow, green line respectively - right panel), surprise is increasingly non-significant as the number of leaves rises (the black, dashed line corresponds to the value $S_{th}=0.05$). These findings point out the existence of a region of the parameter space where surprise misinterprets the planet partition.}
\label{fig3}
\end{figure*}

\begin{equation}
S_\parallel\equiv\sum_{i\geq l_c^*}\sum_{j\geq l_{cp}^*}\frac{\binom{V_c}{i} \binom{V_{cp}}{j} \binom{V-(V_c+V_{cp})}{L-(i+j)}}{\binom{V}{L}}
\label{eq:bs}
\end{equation}
and that we will refer to as to the \emph{bimodular surprise}. The presence of three different binomial coefficients allows three different kinds of links to be accounted for. From a technical point of view, $S_\parallel$ is a p-value computed on a multivariate hypergeometric distribution describing the probability of $i+j$ successes in $L$ draws (without replacement), from a finite population of size $V$ that contains exactly $V_c$ objects with a first specific feature \emph{and} $V_{cp}$ objects with a second specific feature, wherein each draw is either a success or a failure. Although $i$ and $j$ are respectively bounded by the values $V_c$ and $V_{cp}$, analogously to the univariate case, $i+j\in[l_c^*+l_{cp}^*,\min\{L,V_c+V_{cp}\}]$.

The index $c$ in eq. \ref{eq:bs} labels the core part and the index $cp$ labels the core-periphery part; whenever considering bipartite networks, the core-periphery portion will be assumed to indicate the inter-layer portion.

\begin{figure*}[t!]
\centering
\includegraphics[width=0.4\textwidth]{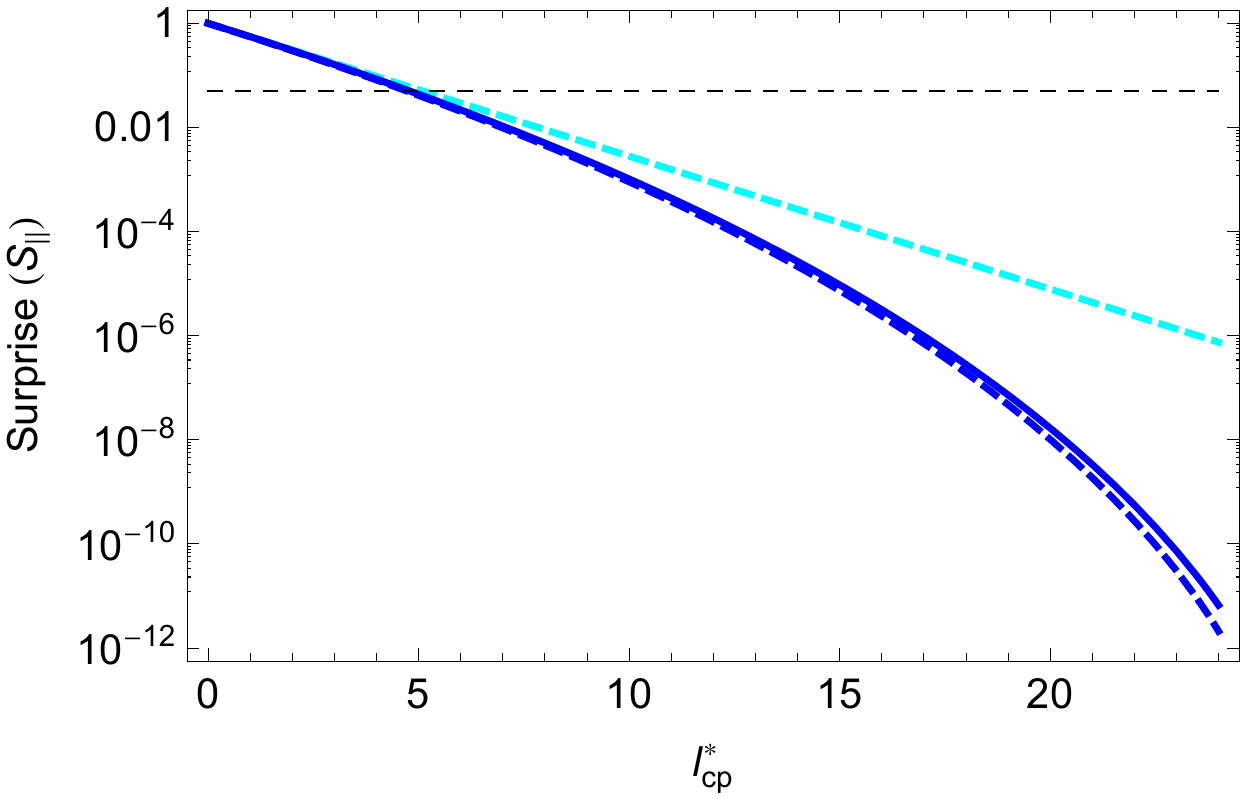}\hspace{3mm}
\includegraphics[width=0.4\textwidth]{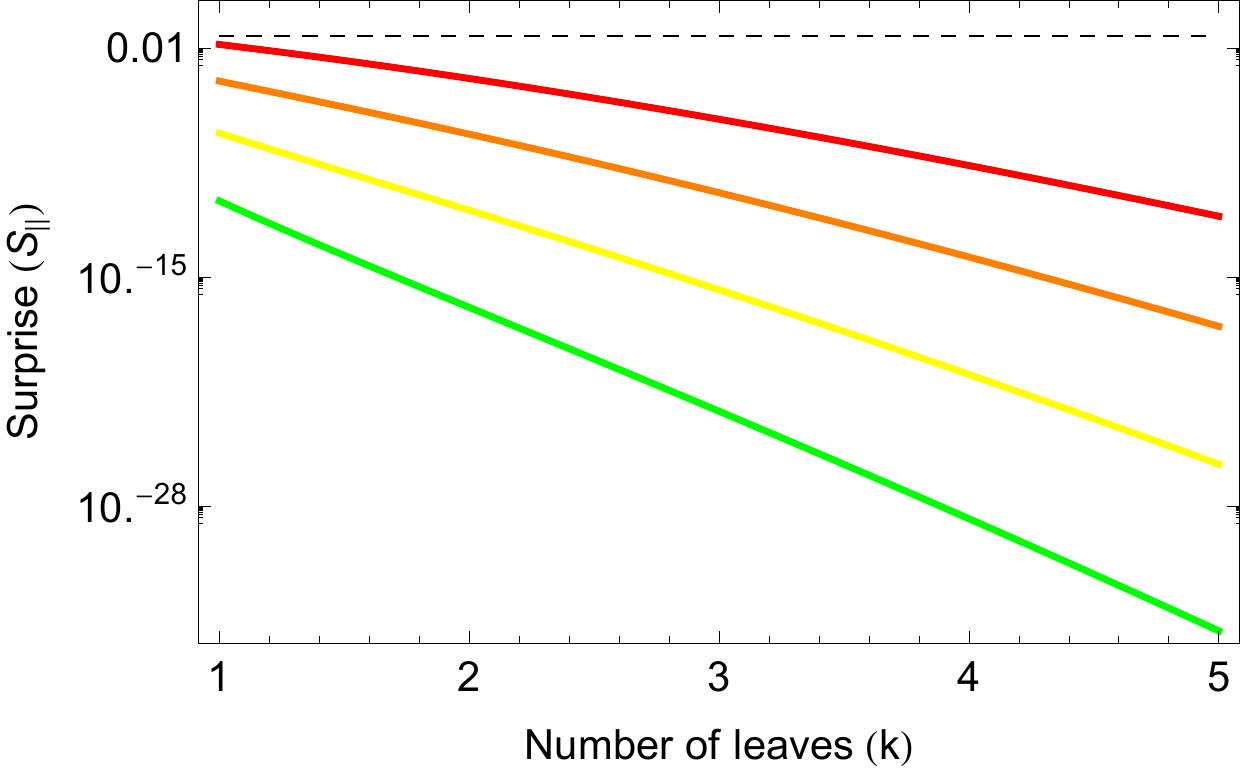}
\caption{Left panel: behavior of $S_\parallel$ as a function of $l_{cp}^*$, for a bipartite network with $N_1=N_2=5$. The blue, solid line corresponds to the (full) expression shown in eq. \ref{eq:pbip}; the blue, dashed line corresponds to the asymptotic expression shown in eq. \ref{eq:pbipapp} and the cyan, dashed line corresponds to its sparse-case approximation. Right panel: behavior of $S_\parallel$ for k-star network configurations (star-like networks are recovered as a particular case, when $k=1$). For a given number of core nodes ($N_1=3,4,5,6$ as indicated by the red, orange, yellow, green line respectively), surprise becomes increasingly significant, as the number of leaves rises. The black, dashed line corresponds to the value $S_{th}=0.05$ in both cases.}
\label{fig4}
\end{figure*}

\paragraph*{Bipartite networks.} Let us now calculate $S_\parallel$ for the bipartite case considered above, defined by the values of parameters $V_c=\frac{N_1(N_1-1)}{2}$ (here, the label $c$ indicates the internal volume of one of the two layers), $V_{cp}=N_1N_2$, $l_c^*=0$ and $l_{cp}^*=L$. The latter condition implies that only the addendum corresponding to $i=0$, $j=l_{cp}^*=L$ survives; thus, our bimodular surprise reads

\begin{equation}
S_\parallel=\frac{\binom{V_{cp}}{l_{cp}^*}}{\binom{V}{l_{cp}^*}}=\frac{\binom{N_1N_2}{l_{cp}^*}}{\binom{(N_1+N_2)(N_1+N_2-1)/2}{l_{cp}^*}}
\label{eq:pbip}
\end{equation}
which \emph{can} be significant, as it should be: in fact, a number of inter-layer links exists above which the observed bipartite structure is significantly denser than its random counterpart (see also fig. \ref{fig4}). Notice that eq. \ref{eq:pbip} can be directly employed to test the significance of any bipartite configuration with no intra-layer links, against the the null hypothesis that such a configuration is compatible with the Random Graph Model: eq. \ref{eq:pbip} shows that, in the considered case, the computation of the searched p-value boils down to calculate the ratio between the number of bipartite networks with $l_{cp}^*$ links and the number of generic configurations with the same number of connections.

\paragraph*{Star-like networks.} Let us now implement our bimodular surprise $S_\parallel$ for star-like configurations. The core portion is identified with the clique of $N_1$ nodes: our parameters, thus, read $V_c=\frac{N_1(N_1-1)}{2}$ and $V_{cp}=N_1^2$. Since, however, $l_c^*=\frac{N_1(N_1-1)}{2}$, the (only) sum indexed by $j$ reduces to the single addendum

\begin{equation}
S_\parallel=\frac{\binom{N_1^2}{N_1}}{{\binom{N_1(2N_1-1)}{N_1(N_1+1)/2}}}
\end{equation}
which is $\simeq 10^{-2}$ for $N_1=3$ and decreases (the corresponding partition, thus, becomes more and more significant) as $N_1$ increases. Notice that the traditional surprise would identify a community structure - with each peripherical node counted as a community on its own - with a comparable significance (see also fig. \ref{fig3}): $S_\parallel$, however, is able to recover the ground-truth structure of the observed network.

\paragraph*{k-star networks.} Analogously, in the k-star case our parameters read $V=\frac{(N_1+kN_1)(N_1+kN_1-1)}{2}$, $V_c=\frac{N_1(N_1-1)}{2}$, $V_{cp}=kN_1^2$, $l_c^*=\frac{N_1(N_1-1)}{2}$ and $l_{cp}^*=kN_1$. Again, thus, the (only) sum indexed by $j$ reduces to just one addendum, i.e.

\begin{equation}
S_\parallel=\frac{\binom{kN_1^2}{kN_1}}{{\binom{(N_1+kN_1)(N_1+kN_1-1)/2}{N_1(N_1-1)/2+kN_1}}}
\label{eq:ks}
\end{equation}
whose behavior is shown in fig. \ref{fig4}: briefly speaking, both in case the number $N_1$ of core nodes rises, while keeping the number of leaves fixed, and the number $k$ of leaves rises, while keeping the number of core nodes fixed, the bimodular surprise becomes increasingly significant, always recovering the ground-truth partition.

\subsection*{Asymptotic results}

The presence of binomial coefficients in the definition of $S_\parallel$ may cause its explicit computation to be demanding from a purely numerical point of view. This subsection is devoted to derive some asymptotic results, in order to speed up the computation of $S_\parallel$. Similar calculations for what concerns the traditional surprise have been carried out in \cite{Traag2015}.

Let us start by considering eq. \ref{eq:pbip}. By Stirling expanding the binomial coefficients appearing in it, one obtains the expression

\begin{equation}
S_\parallel=\frac{\binom{V_{cp}}{l_{cp}^*}}{\binom{V}{l_{cp}^*}}\simeq\frac{p^{l_{cp}^*}(1-p)^{V-l_{cp}^*}}{p_{cp}^{l_{cp}^*}(1-p_{cp})^{V_{cp}-l_{cp}^*}}
\label{eq:pbipapp}
\end{equation}
having defined $p\equiv\frac{l_{cp}^*}{V}$ and $p_{cp}\equiv\frac{l_{cp}^*}{V_{cp}}$ (see the Appendix for the details of the calculations). The expression above makes it explicit that a given (bi)partition is statistically significant if its link density, $p_{cp}$, is large enough to let it be distinguishable from a typical configuration of the Random Graph Model, characterized by link density $p$. In the sparse case, i.e. when $p\ll1$ and $p_{cp}\ll1$, eq. \ref{eq:pbipapp} reduces to $S_\parallel\simeq\left(\frac{p}{p_{cp}}\right)^{l_{cp}^*}$.

Let us now move to the core-periphery case and consider partitions satisfying the condition $l_c^*+l_{cp}^*=L<V_c+V_{cp}$: in this case, one can derive the result

\begin{equation}
S_\parallel=\frac{\binom{V_{c}}{l_{c}^*}\binom{V_{cp}}{l_{cp}^*}}{\binom{V}{L}}\simeq\frac{p^L(1-p)^{V-L}}{p_c^{l_c^*}(1-p_c)^{V_c-l_c^*}\cdot p_{cp}^{l_{cp}^*}(1-p_{cp})^{V_{cp}-l_{cp}^*}}
\label{eq:cpapp}
\end{equation}
having defined $p\equiv\frac{L}{V}=\frac{l_c^*+l_{cp}^*}{V}$, $p_c\equiv\frac{l_c^*}{V_c}$ and $p_{cp}\equiv\frac{l_{cp}^*}{V_{cp}}$. Even if interpreting eq. \ref{eq:cpapp} is less straightforward, it is, however, clear that the significance of the observed partition is a consequence of the interplay between the link density of the core and core-periphery regions (the link density of the periphery has been supposed to be zero - see the Appendix for the details of the calculations).

\section*{Results}

Let us now move to analyze some real-world systems: we will employ our novel definition of surprise to understand if the considered networks have a significant bimodular structure (i.e. either bipartite or core-periphery). To this aim, we will search for the (optimal) partition that minimizes $S_\parallel$ by employing a modified version of the PACO algorithm \cite{Nicolini2016} whose pseudocode is explicitly shown in Appendix and a Python version of which is freely available at \url{https://github.com/jeroenvldj/bimodular_surprise}. In what follows we will consider directed as well as undirected networks.

\begin{figure*}[t!]
\centering
\includegraphics[width=\textwidth]{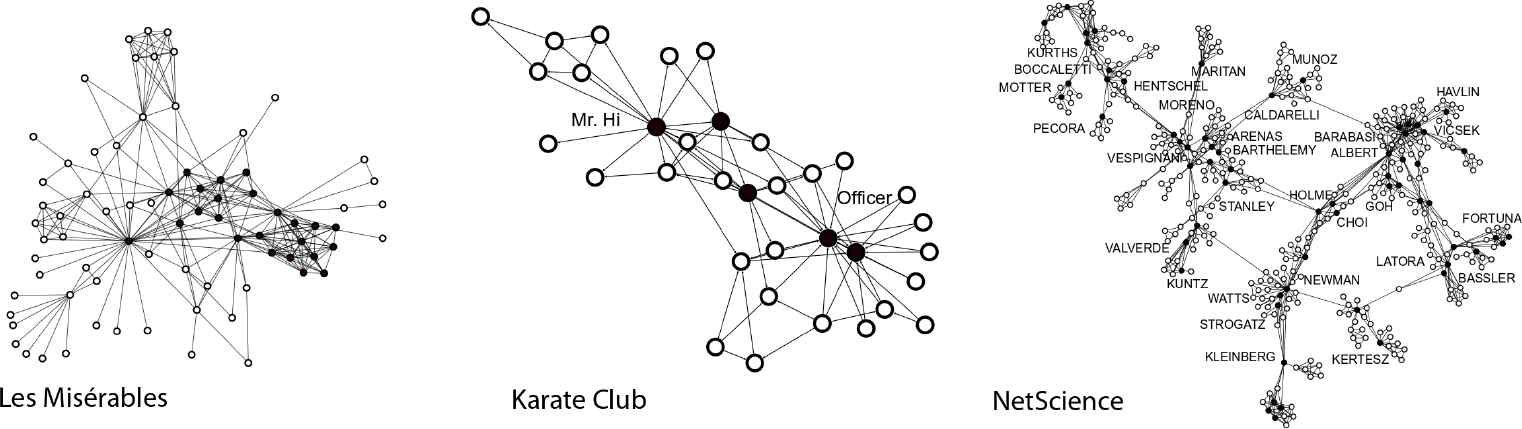}
\caption{Bimodular structure of three real-world social networks (core nodes are drawn in black and periphery nodes are drawn in white). Left panel: core-periphery structure of the network of relationships among ``Les Miserables'' characters; the main characters (e.g. Valjean, Javert, Cosette, Marius) belong to the core. Middle panel: core-periphery structure of the Zachary Karate Club; while the two masters (plus some close disciples) belong to the core, the remaining disciples create a periphery around them, shaping a configuration that is reminescent of the Borgatti \& Everett ideal structure \cite{Borgatti2000}. Right panel: core-periphery structure of the NetSci co-authorship network; while the senior scientists belong to the core - although sharing few direct connections - younger colleagues/students belong to node-specific peripheries connected to the former ones.}
\label{fig5}
\end{figure*}

\paragraph*{Social networks.} Let us start our analysis by considering a number of undirected social networks (see fig. \ref{fig5}). As a first example, let us consider the Zachary Karate Club. Although the latter is commonly employed as a benchmark for community detection, it is also characterized by a clear bimodular structure whose core nodes are represented by the masters, their close disciples and a fifth node ``bridging'' the two masters. Upon looking at the subgraphs constituted by the masters' ego-networks, almost ideal (i.e. \emph{\`a la Borgatti}) core-periphery networks are observable.

A similar comment can be done when considering the network of relationships among the characters of ``Les Miserables'': the main characters (e.g. Valjean, Javert, Cosette, Marius) belong to the core, while the large number of secondary characters linked to them constitute the periphery of such a network (see, for example, the nodes linked to Valjean); intuitively, again, core nodes are very inter-connected while the link density of the periphery is very low. As for the Zachary Karate Club network, there seem to be (core) nodes bridging two dense core subsets.

Let us now consider the (connected component of the) NetSci co-authorship network \cite{netsci}. A core-periphery structure is, again, recovered (although the core is not very dense) where core nodes represent senior scientists (e.g. Stanely, Barabasi, Watts, Kertesz) and periphery nodes represent younger colleagues, students, etc. It is interesting to observe that the senior scientists share relatively few direct connections, while being connected to a plethora of younger collaborators; even more so, the structure of the co-authorship network seems to reflect the structure of the underlying collaboration network, with each research group seemingly being quite separated from the others.

\begin{figure*}[t!]
\centering
\includegraphics[width=0.47\textwidth]{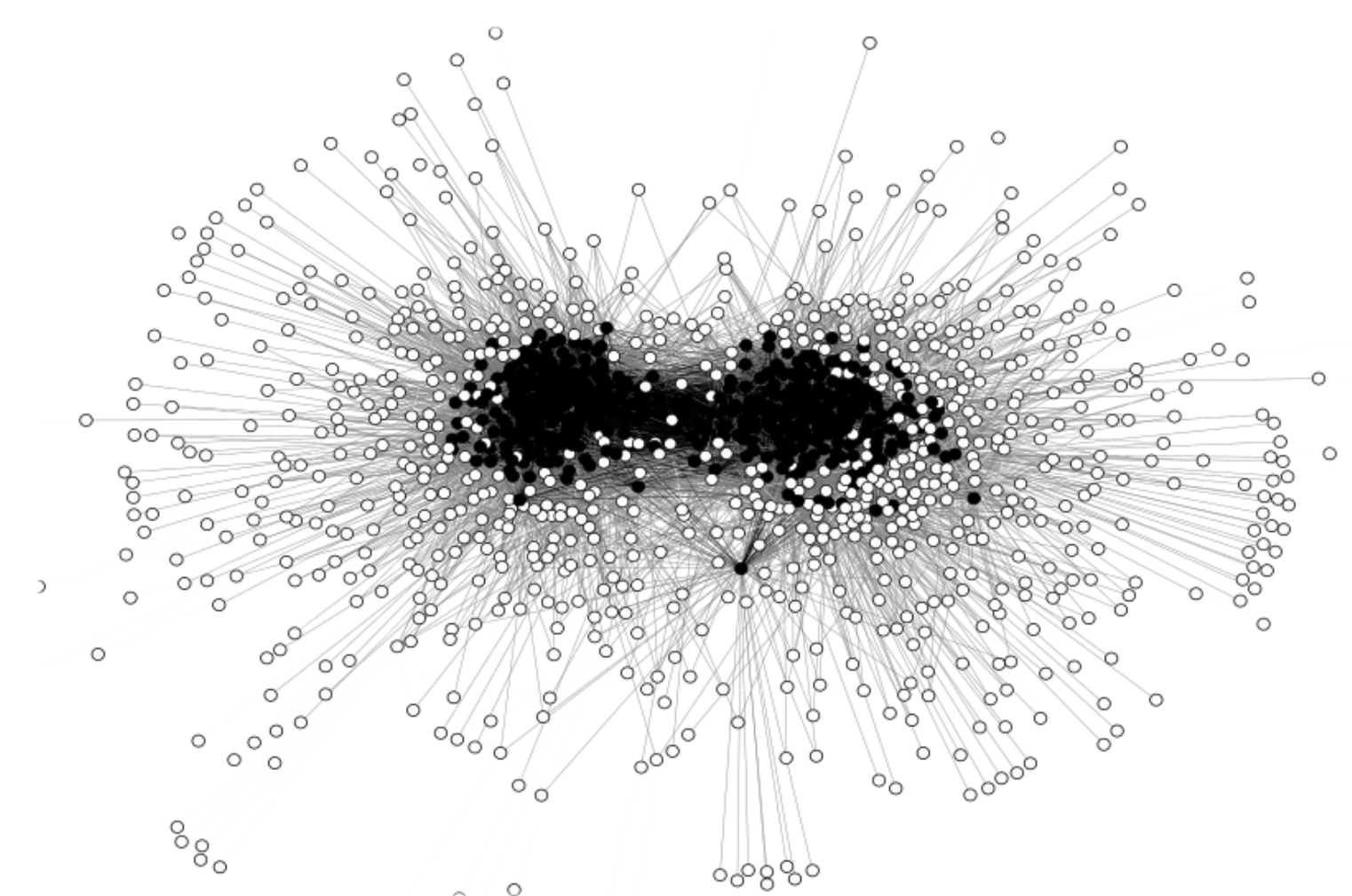}\hspace{3mm}
\includegraphics[width=0.47\textwidth]{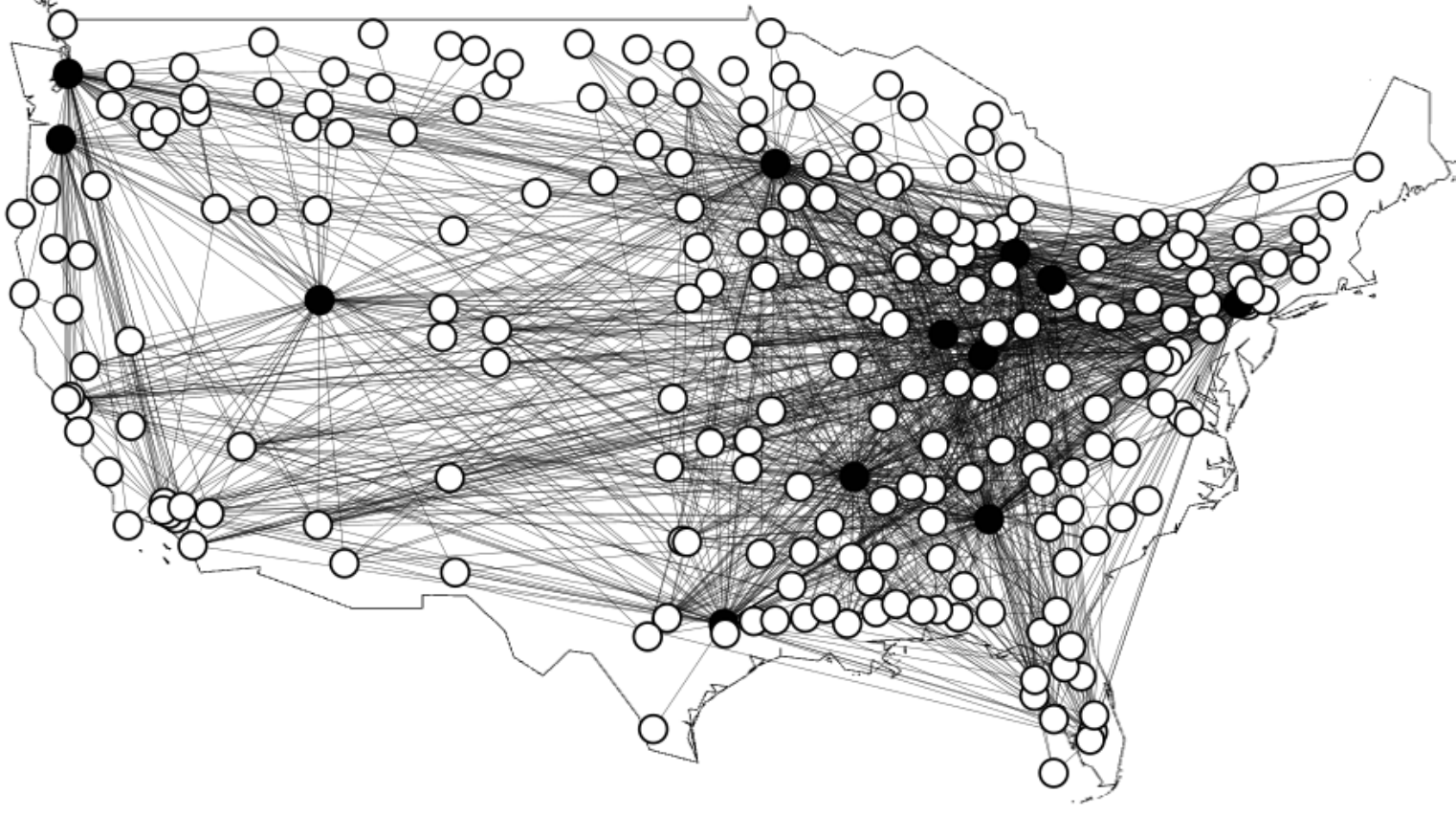}
\caption{Left panel: core-periphery structure of US political blogs \cite{Adamic2005}: a core of the most influential blogs (be they republican or democratic), surrounded by a periphery of loosely connected, less important blogs is clearly visible. Notice that blogs are grouped independently from their political orientation. Right panel: core-periphery structure of US airports. As for the NetSci co-authorship network, each core airport seems to be surrounded by a quite large number of periphery airports, sharing relatively few connections between themselves.}
\label{fig6}
\end{figure*}

A fourth social network is the one showing the relationships between US political blogs \cite{Adamic2005}. Any two blogs are linked if one of the two references the other. As shown in fig. \ref{fig6}, a core of the most influential blogs (be they republican or democratic), surrounded by a periphery of loosely connected, less important blogs is clearly visible. Differently from the community structure that shows republican blogs and democratic blogs as belonging to different groups \cite{Karrer2010}, our core-periphery structure highlights a different organizing principle, based on the blogs overall importance irrespectively from their political orientation. Interestingly enough, the bimodular surprise value indicates that the core-periphery structure is more significant than the traditional republicans VS democrats community structure.

\begin{figure*}[t!]
\centering
\includegraphics[width=\textwidth]{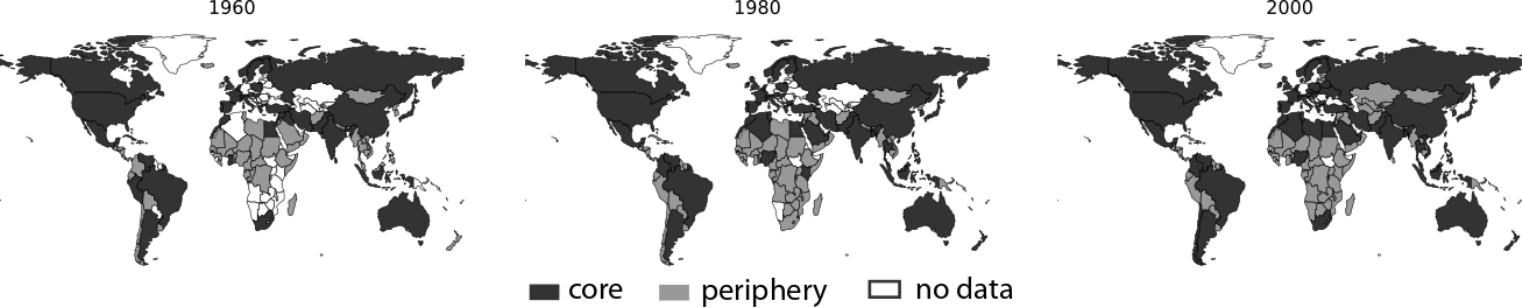}
\caption{Core-periphery structure of the World Trade Web (black: core nodes; gray: periphery nodes). Loosely speaking, while the richest and several developing countries are found to belong to the core, the poorest nations belong to the periphery (e.g. several African nations, throughout our dataset). Notice that core size increases with time: apparently, thus, the system becomes increasingly integrated, confirming a result found in \cite{DeJeude2018}, where it was shown that the size of the WTW strongly connected component increases with time as well.}
\label{fig7}
\end{figure*}

\paragraph*{Economic networks.} Let us now consider an economic network, i.e. the directed representation of the World Trade Web (WTW) in the years 1950-2000: as usual, nodes are world countries and links are trade relationships (i.e. exports, imports) between them. Upon running our bimodular surprise optimization we find a clear core-periphery structure with the core including the richest countries and several developing nations and the periphery including some of the poorest nations (e.g. several African nations throughout our dataset - see also fig. \ref{fig7} where only the years 1960, 1980 and 2000 are shown).

We also observe an interesting dynamics, causing the core size to rise (it represents the $\simeq30\%$ of nodes in 1992 and the $\simeq60\%$ of nodes in 2002) and progressively include countries previously belonging to the periphery. Such a dynamics - that can be interpreted as a signal of ongoing integration - confirms the results found in \cite{DeJeude2018}, where it was shown that the size of the WTW strongly connected component (SCC) increases with time as well. Although the SCC and the core portion of the World Trade Web do not perfectly overlap, many similarities between the two structures are indeed observable.

\paragraph*{Financial networks.} Let us now consider a financial network, i.e. e-MID, the electronic Italian Interbank Market. Here, we compare two different datasets: the first one collects the 2005-2010 interbank transactions during the so-called maintenance periods \cite{Hatzopoulos2015}; the second one collects interbank transactions on a daily basis from 1999 to 2012 \cite{Barucca2015,Barucca2018}. The main difference between the two datasets lies in their level of aggregation: notice, in fact, that the first one basically collects data on a monthly basis.

\begin{figure}[t!]
\centering
\includegraphics[width=0.41\textwidth]{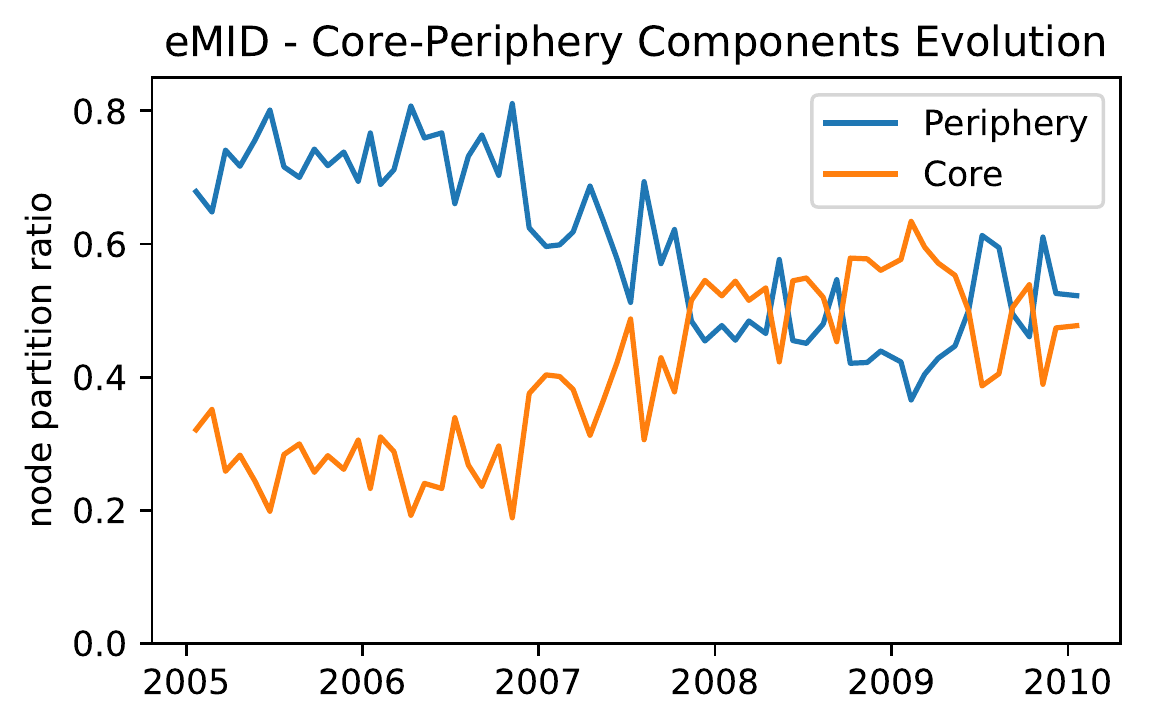}\vspace{3mm}\\
\includegraphics[width=0.41\textwidth]{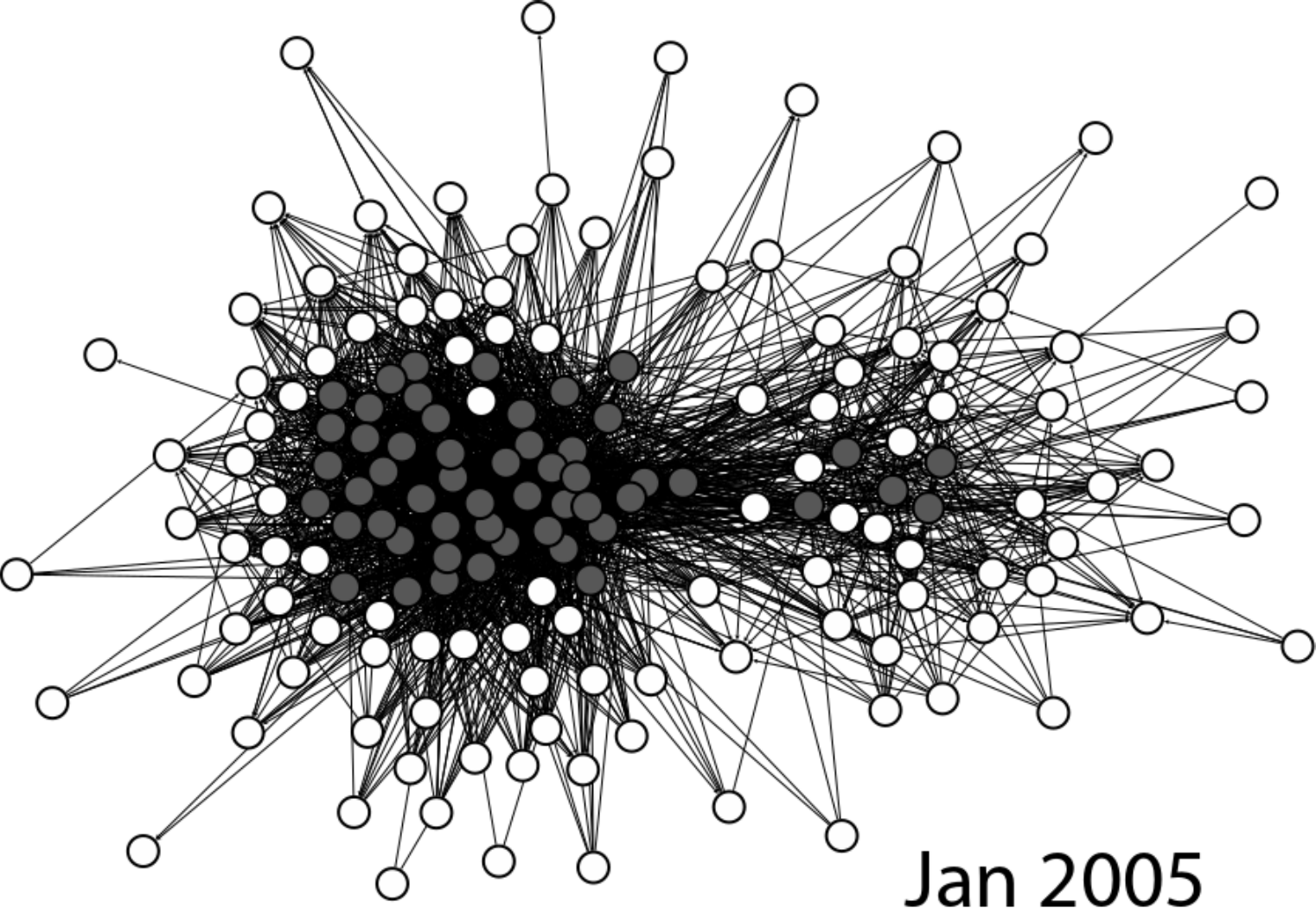}\vspace{3mm}\\
\includegraphics[width=0.41\textwidth]{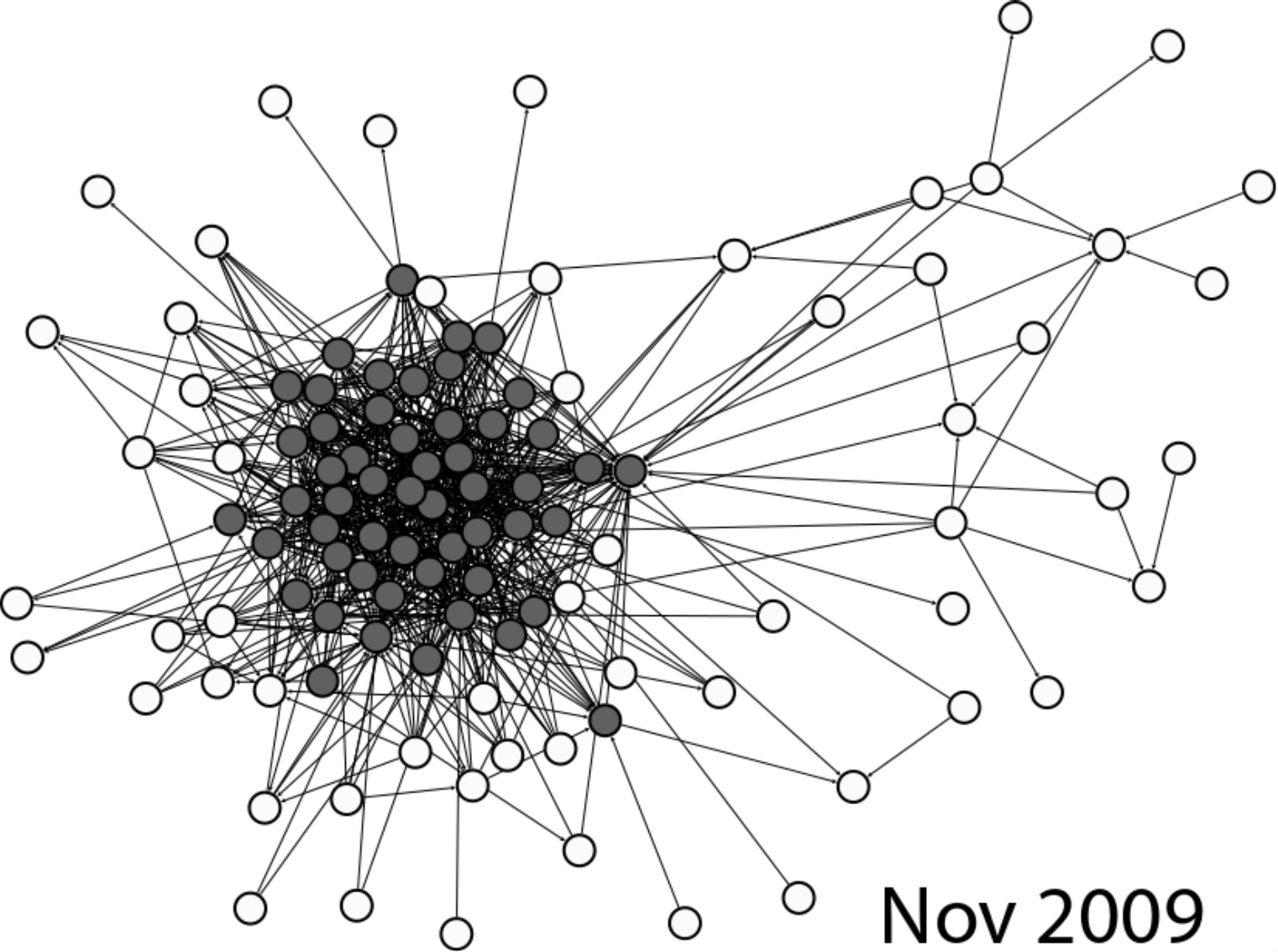}
\caption{Core-periphery structure of e-MID maintenance periods (gray: core nodes; white: periphery nodes). After an initial period of two years characterized by an approximately constant value of the core and periphery size, a structural change takes place in 2007 and the percentage of nodes belonging to the core steadily rises until 2008. Afterwards, an equilibrium seems to be re-established. This may be due to a decrease in the total number of nodes which, however, does not affect the type of banks belonging to the core (italian banks) and to the periphery (foreign banks). Networks are directed but we have omitted the link directionality for the sake of readability.}
\label{fig8}
\end{figure}

Let us start by analyzing the first dataset. As fig. \ref{fig8} shows, its structure undergoes an interesting evolution: after an initial period of two years, where a large periphery of loosely connected nodes ($\simeq 70\%$) exists, a transient period of one year (i.e. 2007) during which the percentage of nodes belonging to the core rises, is visible. Afterwards, an equilibrium situation seems to be re-established with the percentage of core and periphery nodes basically coinciding. Even if the total number of banks registered in the dataset steadily decreases after 2007, this doesn't seem to affect the type of banks belonging to the core and to the periphery, i.e. Italian and foreign banks, respectively.

\begin{figure}[t!]
\centering
\includegraphics[width=0.45\textwidth]{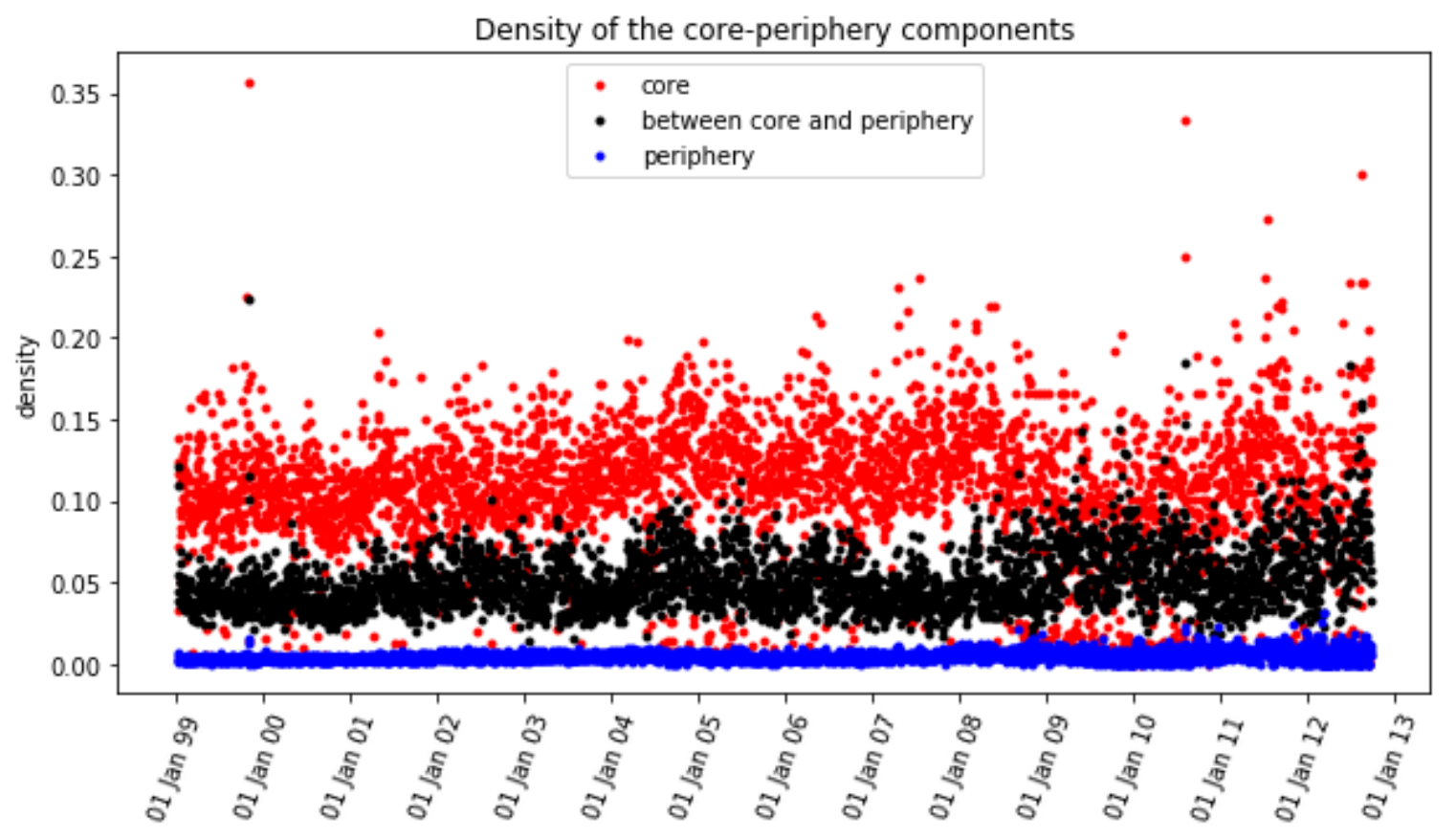}\vspace{3mm}\\
\includegraphics[width=0.3\textwidth]{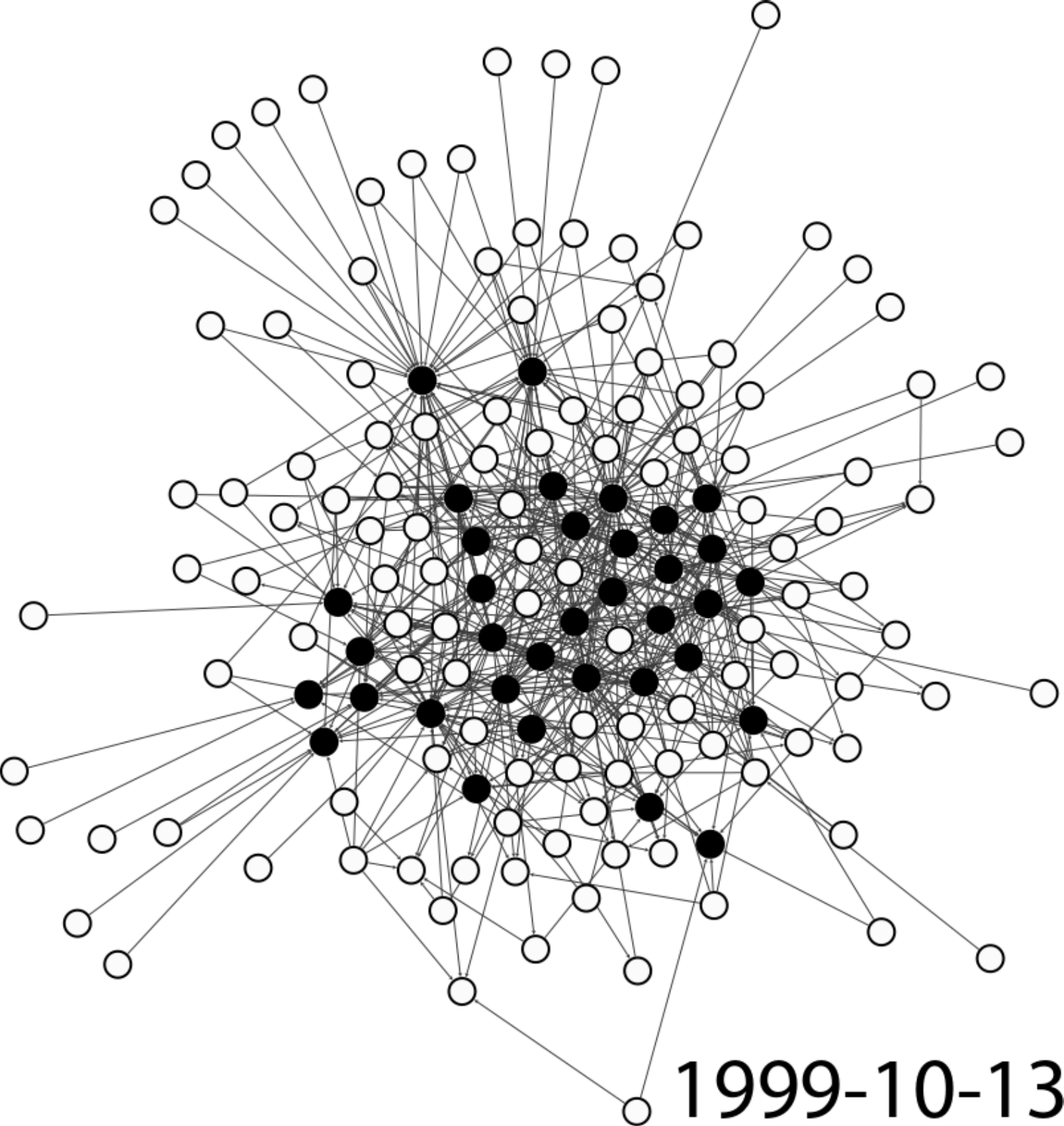}\vspace{3mm}\\
\includegraphics[width=0.355\textwidth]{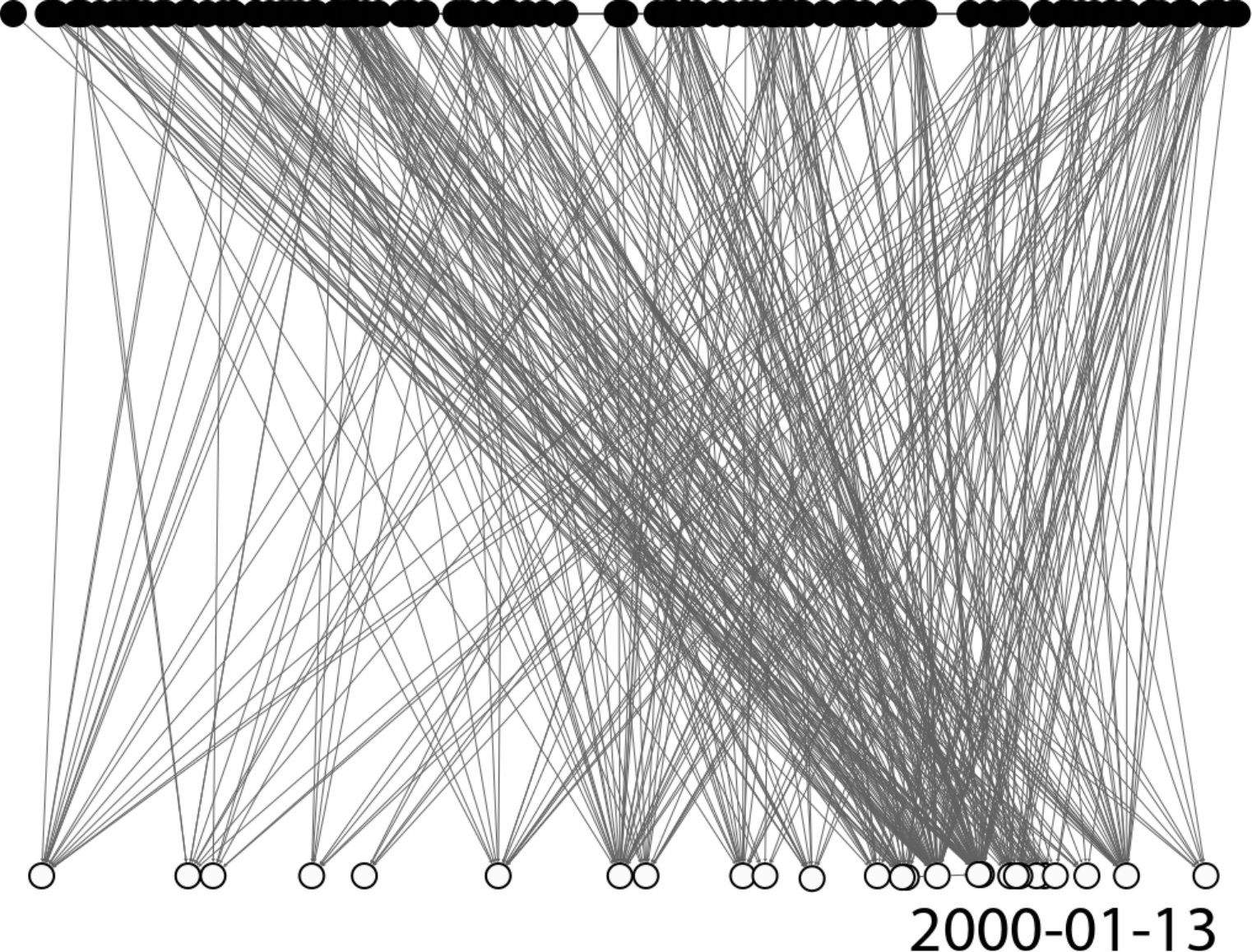}
\caption{Mesoscale structure of e-MID daily data. Although for the vast majority of snapshots a core-periphery structure seems to better represent the e-MID network, the number of times a bipartite structure is observed increases after 2008. Middle and bottom panels explicitly show two different snapshots of e-MID: the first one is characterized by the chain of inequalities $c_p<c_{cp}<c_c$; the second one, instead, shows a configuration for which the values $c_p\simeq c_c<c_{cp}$ are observed, indicating the presence of a bipartite structure (when referring to bipartite structures, the label $cp$ is assumed to indicate the inter-layer portion).  Networks are directed but we have omitted the link directionality for the sake of readability.}
\label{fig9}
\end{figure}

Let us now move to the analysis of the second dataset. As fig. \ref{fig9} shows, the analysis of the link density of the portions in which $S_\parallel$ partitions the network reveals that, overall, a core-periphery structure seems to characterize the daily data better than a bipartite structure. This picture, however, seems to be less correct from 2008 on: as the last portion of the first panel of fig. \ref{fig9} shows, a bipartite structures occour more often than a core-periphery structure during this period. Two snapshots of the network are also explicitly shown, illustrating the values of link density characterizing the different network portions.

\paragraph*{Other kinds of networks.} As a last example, let us consider the US airports network (see fig. \ref{fig6}). Examples of core airports are the ones of New York, Indianapolis, Salt Lake City, Seattle, etc. The periphery airports are preferentially attached to the core ones. This system shares interesting similarities with the NetSci co-authorship network: each core airport, in fact, seems to be surrounded by a quite large number of periphery airports, sharing few internal connections.

\section*{Discussion}

It is hard to underestimate the importance of the presence of bimodular mesoscale structures in real-world networks: while the authors in \cite{Peixoto2012} show that the most robust topology against random failures is the core-periphery one, understanding the relationship between a given node systemicness and its coreness is of paramount importance in finance \cite{Luu2018}. In the same field, a core-periphery structure is believed to reflect the ``essential'' function of banks: the core ones tie the periphery ones into a single market through their intermediation activity \cite{Craig2010}. On the other hand, a bipartite structure would reflect the absence of intermediation, i.e. a market displaying preferential trading \cite{Barucca2015}.

In this paper we have proposed a novel measure for bimodular mesoscale structures detection. To this aim, we have adopted a surprise-like score function, by considering the multivariate version of the quantity proposed in \cite{Nicolini2016}. Employing this kind of quantities means implementing a bottom-up approach, i.e. letting the modular structure to be extrapolated from the data and not imposed \emph{a priori} as in previous approaches \cite{Borgatti2000,Barucca2018}.

Most importantly, such a comparison is based on a properly-defined null model, allowing the significance of a given partition to be quantifiable via a p-value. As for the traditional surprise, the reference model is the (Directed) Random Graph Model that constrains the total number of observed connections, while randomizing everything else. The choice of employing such a benchmark is dictated by a number of recent results, pointing out that several mesoscale structures of interest (e.g. the core-periphery one, the bow-tie one, etc.) are actually compatible with - and hence undetectable under - a null model constraining the entire degree sequence(s) \cite{IntVeld2014,Kojaku2018}.

While solving the problem of consistently comparing an observed structure with a ``random'' model of it, our approach also solves a second drawback affecting the methods in \cite{Craig2010,Borgatti2000} and pointed out in \cite{Kojaku2018}: ideal structures as the ones searched by algorithms \emph{\`a la Borgatti} are very reliant on the nodes degree, with the core often composed of just the nodes with the largest number of neighbors. This is not necessarily true when a benchmark is adopted for comparison \cite{Zhang2014}: as previously discussed, the significance of a given partition detected by surprise results from the interplay between the link density values of the different network areas.

This also sheds light on the relationship between apparently conflicting structures co-existing within the same network configuration: generally speaking, traditional and bimodular surprise optimization should be considered complementary - rather than mutually exclusive - steps of a more general analysis. As the example of the US political blogs confirms, it is indeed possible that a community structure co-exists with a core-periphery structure; a second, less trivial, example is provided by the World Trade Web, whose community structure has been studied in \cite{Barigozzi2011} but whose significance has, then, been questioned \cite{Piccardi2012}.

As a last comment, we would like to stress that the two approaches to mesoscale structures detection that have been proposed so far - comparing an observed structure with a benchmark \cite{Kojaku2017,Kojaku2018} and searching for the model best fitting a given partition \cite{Zhang2014,Barucca2015,Barucca2018,Karrer2010} - can be supposed to be complementary, since a non-significant structure under a given benchmark is surely more compatible with it. Employing a benchmark, however, provides an advantage, i.e. making the statistical significance of a given structure explicit - something that remains ``implict'' when employing the fitting procedure. In other words, searching for the best fit may push one to enrich a model with an increasing amount of information whose relevance cannot be easily clarified. Such a problem seems to affect all likelihood-based algorithms unless a more refined criterion to judge the goodness of a fit is employed: solutions like the one of adopting criteria like the Akaike Information Criterion \emph{et similia} have been proposed \cite{Burnham2002}.

The present work calls for a generalization to \emph{weighted} mesoscale structures detection, a field where relatively little has been done so far \cite{Nicolini2016b,Fagiolo2008}.

\section*{Appendix}

The computation of binomial coefficients for large graphs can quickly become numerically demanding. In order to simplify the calculations of our bimodular surprise, let us proceed by steps. First, let us Stirling approximating the binomial coefficients:

\begin{widetext}
\begin{eqnarray}
\binom{V_c}{i}&\simeq&\left[(p_c)^i\left(1-p_c\right)^{V_c-i}\right]^{-1},\\
\binom{V_{cp}}{j}&\simeq&\left[(p_{cp})^j\left(1-p_{cp}\right)^{V_{cp}-j}\right]^{-1},\\
\binom{V_p}{L-(i+j)}&\simeq&\left[(p_p)^{L-(i+j)}\left(1-p_p\right)^{V_p-(L-(i+j))}\right]^{-1},\\
\binom{V}{L}&\simeq&\left[p^L\left(1-p\right)^{V-L}\right]^{-1}
\end{eqnarray}
\end{widetext}
having defined $V_p\equiv V-(V_c+V_{cp})$, $p\equiv\frac{L}{V}$, $p_c\equiv\frac{i}{V_c}$, $p_{cp}\equiv\frac{j}{V_{cp}}$, $p_p\equiv\frac{L-(i+j)}{V_p}$. As a second step, let us substitute the expressions above into eq. \ref{eq:bs}:

\begin{widetext}
\begin{equation}
S_\parallel\simeq\sum_{i\geq l_c^*}\sum_{j\geq l_{cp}^*}\left(\frac{p}{p_p}\right)^L\left(\frac{1-p}{1-p_p}\right)^{V-L}\left(\frac{p_p}{p_c}\right)^i\left(\frac{1-p_p}{1-p_c}\right)^{V_c-i}\left(\frac{p_p}{p_{cp}}\right)^j\left(\frac{1-p_p}{1-p_{cp}}\right)^{V_{cp}-j};
\label{eq:bs2}
\end{equation}
\end{widetext}
in order to obtain a more explicit expression, let us limit ourselves to consider the leading term of the summation in eq. \ref{eq:bs2} that is readily obtained upon substituting $i$ with $l_c^*$ and $j$ with $l_{cp}^*$:

\begin{widetext}
\begin{equation}
S_\parallel\simeq\left(\frac{p}{p_p}\right)^L\left(\frac{1-p}{1-p_p}\right)^{V-L}\left(\frac{p_p}{p_c}\right)^{l_c^*}\left(\frac{1-p_p}{1-p_c}\right)^{V_c-l_c^*}\left(\frac{p_p}{p_{cp}}\right)^{l_{cp}^*}\left(\frac{1-p_p}{1-p_{cp}}\right)^{V_{cp}-l_{cp}^*}
\label{eq:bs3}
\end{equation}
\end{widetext}
where, now, $p_c\equiv\frac{l_c^*}{V_c}$, $p_{cp}\equiv\frac{l_{cp}^*}{V_{cp}}$, $p_p\equiv\frac{L-(l_c^*+l_{cp}^*)}{V_p}$. Notice that eq. \ref{eq:bs3} can be employed to detect both bipartite and core-periphery structures, upon identifying the core and the periphery portions as the network portions \emph{within} layers. Eq. \ref{eq:bs3} already makes intuitively clear that our bimodular surprise is likely to be significant either when $p_c\simeq p_p$ but $p_{cp}\gg p_c\simeq p_p$ (i.e. in the case of bipartite networks - notice that eq. \ref{eq:pbipapp} is recovered when $L=l_{cp}^*$) or when $p_c\gg p_p$ and $p_{cp}\gg p_p$ (i.e. in the core-periphery case - notice that eq. \ref{eq:cpapp} is recovered when $L=l_c^*+l_{cp}^*<V_c+V_{cp}$).

\begin{figure}[t!]
\centering
\includegraphics[width=0.49\textwidth]{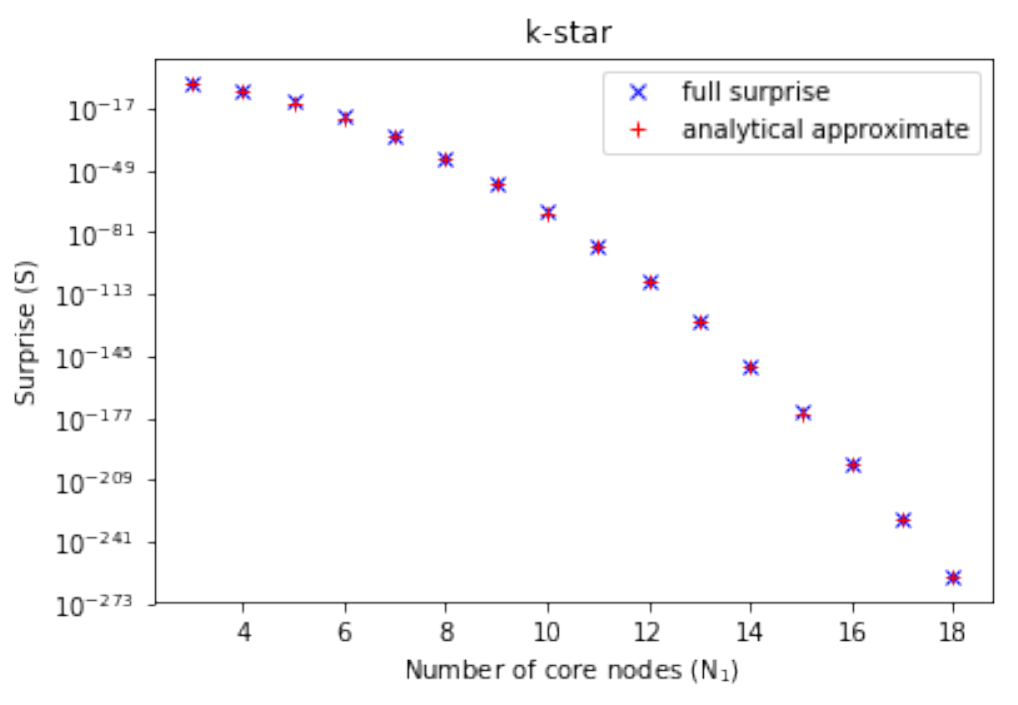}
\caption{Numerical check of the validity of the approximation shown in eq. \ref{eq:bs3}, computed for k-star networks and compared to the full expression in eq. \ref{eq:bs}.}
\label{fig10}
\end{figure}

A numerical check of the validity of the proposed approximation is shown in fig. \ref{fig10}, where it is computed for k-star networks and compared to the full expression in eq. \ref{eq:bs}. We explicitly notice, however, that the approximation shown in eq. \ref{eq:bs3} is recommended for analysing small networks; when large networks are considered, the full expression in eq. \ref{eq:bs} is, instead, recommended.

\subsection*{Numerical surprise optimization:\\a modified PACO algorithm}

In what follows we show a modified version of the PACO (PArtitioning Cost Optimization) algorithm pseudocode \cite{Nicolini2016} by running which the partition that minimizes surprise can be found. The PACO algorithm implements an approach that is heuristic in nature since an exahustive search of all possible partitions is not feasible when dealing with large graphs.

The idea is that of assigning every node to either one of two subsets - nterpretable as the core and the periphery or the layers of a bipartite graph - by running a greedy process that takes as input pairs of nodes connected by an edge and evaluates whether those two nodes should belong to the same subset or not: the choice that minimizes the surprise is the one that is actually implemented. In order to speed up the calculations, the original PACO algorithm takes edges that are previously sorted according to their decreasing value of Jaccard index. Since the latter quantifies the fraction of common neighbours of the two connected nodes, nodes pairs with larger Jaccard index are also the ones most likely to be assigned to the same subset (e.g. a community).

Since for pure bipartite graphs the Jaccard index - as defined above - is zero for all edges (nodes connected by an edge always lie on different layers) we need to modify the score according to which we sort the edges. In our modified version of the PACO algorithm we sort links according to the number of z-motifs they belong to, the latter being defined as $z_{i\alpha}=\sum_{\beta,j}a_{i\beta}a_{i\alpha}a_{\alpha j}$: in other words, we evaluate the number of times a generic link is the ``middle'' one of a path whose length is 3. As with the original PACO algorithm, we progressively consider all edges, sorted as described above, evaluating wheter the linked pairs should belong to the same subset or not.

As a final step, we consider a number of random reassignments of nodes with the aim of preventing the possibility of getting stuck in a local minimum (a random move consists of selecting 3 random nodes belonging to the same group and evaluating if assigning them to different subsets would further minimize surprise).

The algorithm described above performs quite well in finding the global minimum of surprise on a range of different configurations we have tested. When considering low-density bipartite graphs, however, the algorithm does not always succeed in reaching the global minimum.\\

\begin{algorithmic}[1]
\Function{CalculateAndUpdateSurprise}{$C,C'$}
    \State $S\gets calculateSurprise(C)$
    \State $S'\gets calculateSurprise(C')$
    \If{$S'<S$}
        \State $C\gets C'$
        \State $S\gets S'$
    \EndIf \\
    \Return $C$
\EndFunction
\State
\State $C \gets $ array of length $N$ randomly initialized with binary entries (0 or 1);
\State $E \gets $ sorted edges in decreasing order;
\For{edge $(u,v)\in E$}
    \State $C'\gets C$
    \If{$C'[u] \neq C[v]$}
        \State $C'[u]\gets C[v]$
        \State $C\gets$\Call{CalculateAndUpdateSurprise}{$C,C'$}
    \Else
        \State $C'[u]\gets 1-C[v]$
        \State $C\gets$\Call{CalculateAndUpdateSurprise}{$C,C'$}
    \EndIf
\State $\Rightarrow$ randomly switch node membership for $n=3$ nodes in the same partition and accept move if $S_\parallel$ decreases;
\EndFor
\State{ $\Rightarrow$ repeat several times the for-loop to improve the chance of finding the optimal partition.}
\end{algorithmic}

\section*{Acknowledgements}

This work was supported by the EU projects CoeGSS (grant num. 676547), DOLFINS (grant num. 640772), MULTIPLEX (grant num. 317532), Openmaker (grant num. 687941), SoBigData (grant num. 654024).

\section*{Authors Contributions}

JLJ and TS developed the method. JLJ performed the analysis. JLJ, GC and TS wrote the manuscript. All authors reviewed and approved the manuscript.

\section*{Additional Information}

The authors declare no competing financial interests.

\end{document}